\documentclass[conference]{IEEEtran}
\IEEEoverridecommandlockouts
\usepackage{soul}
\usepackage{cite}
\usepackage{amsmath,amssymb,amsfonts}
\usepackage{algorithmic}
\usepackage{graphicx}
\usepackage{textcomp}
\usepackage{xcolor}
\usepackage{comment}
\usepackage{array}
\usepackage{multirow}
\usepackage{caption}
\usepackage{subcaption}
\usepackage{graphicx}
\usepackage{url}
\usepackage{tikz}
\newcommand*\circled[1]{\tikz[baseline=(char.base)]{
            \node[shape=circle,draw,inner sep=1pt,font=\sffamily\footnotesize] (char) {\textbf{#1}};}}
\mathchardef\mhyphen="2D
\def\BibTeX{{\rm B\kern-.05em{\sc i\kern-.025em b}\kern-.08em
    T\kern-.1667em\lower.7ex\hbox{E}\kern-.125emX}}
\begin{document}

\title{Automatic Mapping of Unstructured Cyber Threat Intelligence: An Experimental Study\\
{\normalsize \textbf{(Practical Experience Report)}}}

\author{
    \IEEEauthorblockN{Vittorio Orbinato\IEEEauthorrefmark{1}, Mariarosaria Barbaraci\IEEEauthorrefmark{2}, Roberto Natella\IEEEauthorrefmark{1}, Domenico Cotroneo\IEEEauthorrefmark{1}}
    \IEEEauthorblockA{\IEEEauthorrefmark{1}\textit{DIETI}, Università degli Studi di Napoli Federico II, Naples, Italy
    \\\{vittorio.orbinato, roberto.natella, cotroneo\}@unina.it}
    \IEEEauthorblockA{\IEEEauthorrefmark{2}University of Bern, Bern, Switzerland
    \\mariarosaria.barbaraci@unibe.ch}
}

\maketitle

\begin{abstract}
Proactive approaches to security, such as adversary emulation, leverage information about threat actors and their techniques (Cyber Threat Intelligence, CTI). However, most CTI still comes in unstructured forms (i.e., natural language), such as incident reports and leaked documents. To support proactive security efforts, we present an experimental study on the automatic classification of unstructured CTI into attack techniques using machine learning (ML). We contribute with two new datasets for CTI analysis, and we evaluate several ML models, including both traditional and deep learning-based ones. We present several lessons learned about how ML can perform at this task, which classifiers perform best and under which conditions, which are the main causes of classification errors, and the challenges ahead for CTI analysis.
\end{abstract}


\section{Introduction}
\label{sec:Introduction}


Addressing security issues of IT systems has become of paramount importance for all kinds of businesses. Traditional penetration testing alone, consisting mostly in identifying and exploiting vulnerabilities, is becoming increasingly ineffective \cite{Applebaum2016IntelligentAR}. Its exclusive focus on the first steps of the cyber kill chain \cite{killchain} prevents analysts from deepening into post-compromise scenarios, by analyzing what might happen in the post-exploitation phases of an attack. In order to strengthen security, organizations are considering ``offensive security'' strategies based on adversary emulation \cite{Applebaum2016IntelligentAR}, i.e., a proactive approach that emulates attackers within an IT system for training and evaluation purposes. Unfortunately, this approach comes with a high cost, in terms of specialized personnel (``red teams'') needed to emulate attackers, which limits its adoption.


In recent years, adversary emulation tools have been emerging as a promising solution to make  security exercises easier \cite{Sok}. These tools automate the execution of individual, low-level malicious actions, such as for credential stealing, lateral movement, data exfiltration, and more \cite{Applebaum2017AnalysisOA}. Despite this support, these tools still need to be programmed to orchestrate multiple actions and to emulate the behaviors of a real attacker. Therefore, adversary emulation exercises need to be guided by \emph{Cyber Threat Intelligence} (CTI), i.e., information on the capabilities and intents of attackers, and on the techniques adopted in their campaigns.

To support adversary emulation and other security activities, the industry is currently investing efforts in standardizing CTI, such as the STIX representation format \cite{STIXMITRE} and the TAXII sharing protocol \cite{TAXIIMITRE}. 
However, most Cyber Threat Intelligence still comes in unstructured forms (i.e., text in natural language), such as incident reports written by security analysts\cite{FIN6FireEye, WizardSpiderCISA, MenuPassSymantec}, and documents leaked by insiders from attacker groups \cite{TheRecordConti, MuddyWaterFreeLibrary, OilRig}. Therefore, converting CTI into a structured form is still a missing step towards supporting adversary emulation with automated tools.

In this work, we present a study on automatically extracting CTI from unstructured sources, using machine learning techniques for Natural Language Processing (NLP). We contribute with \textbf{two new datasets for CTI analysis}, with samples of CTI in natural language text, which we labeled with the corresponding adversarial techniques described in the text. We based the labeling on the well-known MITRE ATT\&CK framework \cite{attack}, which provides an extensive knowledge base and classification scheme of techniques used by most cyber criminal groups. We also present an extensive \textbf{experimental analysis on classifying unstructured CTI} into the corresponding attack techniques, by using both traditional machine learning algorithms, and more recent ones based on deep learning.

From the experimental analysis, we derived some \textbf{lessons learned} about the automatic mapping of unstructured CTI. CTI mapping is a difficult multi-class classification problem, since, at the time of writing, the ATT\&CK framework describes 188 different attack techniques. Moreover, there is an imbalance among the techniques, since some of them occur much more frequently in real attacks. We found that only few algorithms can achieve an accurate classification, and that algorithms based on deep learning can only achieve a competitive accuracy when trained with our new extended dataset. We also qualitatively analyzed classification mistakes, in order to identify the reasons for inaccuracies. We found that applying classifiers on short texts is prone to errors, since descriptions can be ambiguous and can be equally apply to multiple techniques. Finally, we observe that evaluating ML techniques on descriptions of individual techniques can be misleading. Therefore, we analyzed CTI documents as a whole, which span multiple techniques adopted in the same campaign, and which contain more noise due to sentences not strictly related to attack techniques. We found that indeed ML models perform differently in the document-level evaluation compared to the sentence-level evaluation, and that deep learning techniques can achieve better results in this scenario.


The paper is structured as follows: Section \ref{sec:Approach} provides background and a motivating example; Section \ref{sec:Classification} illustrates the experimental process of this work; Section \ref{sec:Results} and \ref{sec:Document} present the experimental results, respectively on sentence-level and document-level datasets; Section \ref{sec:lessons-learned} analyzes the lessons learned; Section~\ref{sec:Related} discusses related work; and Section \ref{sec:Conclusion} draws the conclusion and points out some directions for future work.

\section{Background and Motivating Example}
\label{sec:Approach}

The planning and execution of adversary emulation exercises are expensive activities, in terms of skills and man-hours to invest. They include teaming as well as other activities: scenario definition, monitoring, and results evaluation.
Adversary emulation is composed by a set of activities which can be split into two macro-phases: \textit{Attack modeling} and \textit{Attack emulation}.

\vspace{3pt}
\noindent
\textbf{Attack modeling}. This phase concerns Cyber Threat Intelligence and it gathers information on the threats and attacks to emulate. A significant effort is required to gather these documents, because of their heterogeneous nature and sources. It is then necessary to analyze and systematize this data, so as to map an attack onto a set of operational steps, in order to define an execution flow for the emulation exercise. 
To this goal, many cybersecurity frameworks provide taxonomies of attack tactics, techniques, and procedures (TTPs), with MITRE ATT\&CK \cite{attack} rapidly becoming the \textit{de facto} standard in the cybersecurity landscape. 
All the activities required by this phase are currently performed by human operators, hence the need for highly specialized personnel.

\begin{figure}[htbp]
\centering
\centerline{\includegraphics[width=\columnwidth]{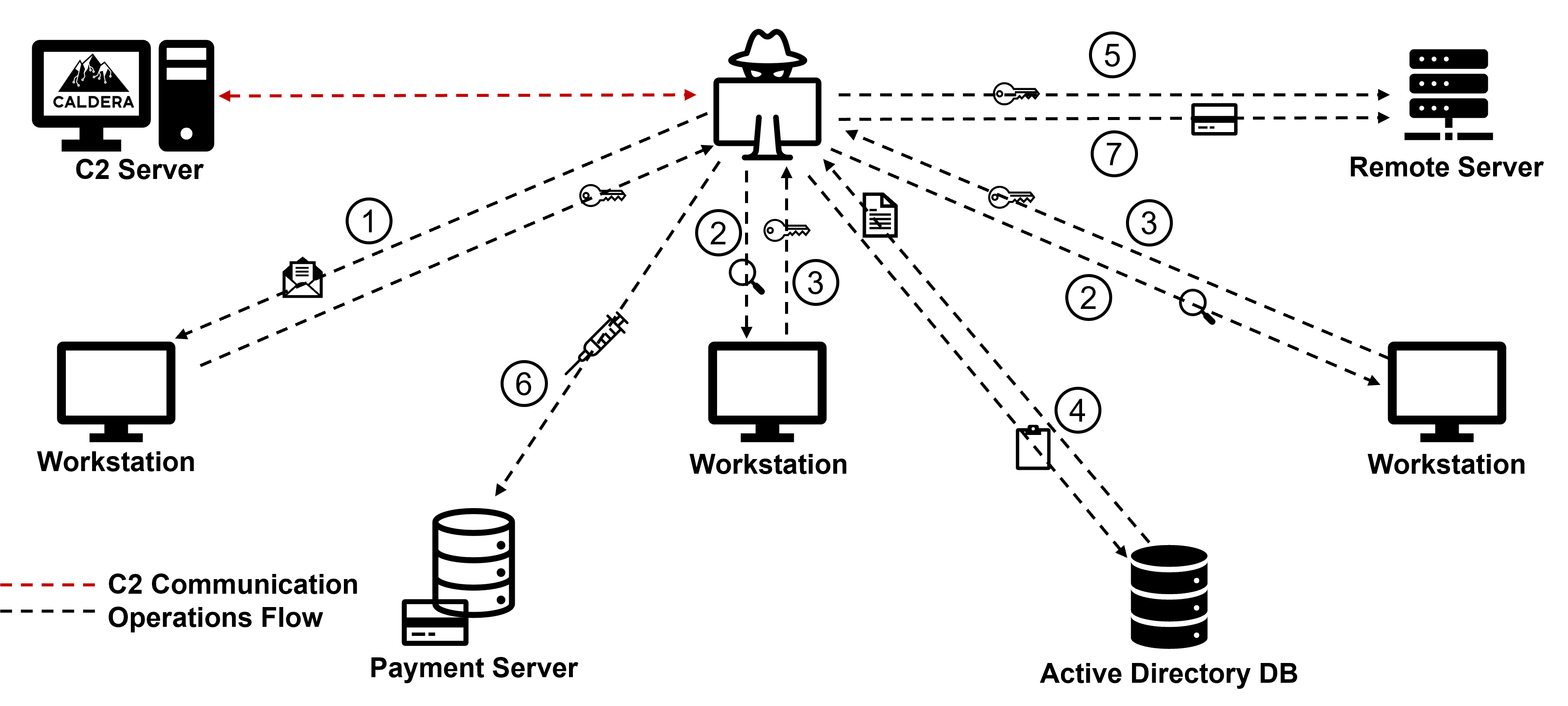}}
\caption{FIN6 Operational Steps.}
\label{fig:FIN6Steps}
\end{figure}

As an example, we will refer to the FIN6 APT \cite{fin6}, analyzed and documented by the Center for Threat-Informed Defense (CTID) \cite{CTID}. FIN6 is a financially motivated cyber-criminal group, whose activities can be traced back to 2015 \cite{fin6summary}. To identify their activities and describe their \textit{modus operandi}, CTID collected and analyzed several threat reports from different organizations, such as VISA \cite{FIN6Visa}, FireEye \cite{FIN6FireEye}, and Security Intelligence \cite{FIN6SecurityIntelligence}. 
FIN6's activities were mapped to MITRE ATT\&CK techniques, and divided into \textit{Enabling} \cite{fin6phase1} and \textit{Operational} \cite{fin6phase2} objectives, whose operational steps are shown in Figure \ref{fig:FIN6Steps}. Their first tactical objective is \emph{Initial Access}, which identifies potential targets to be exploited, by means of legitimate \emph{stolen user credentials} (T1078\footnote{\textbf{Txxxx} indicates a technique from the MITRE ATT\&CK framework} \cite{T1078}) and \emph{spearfishing} (T1566 \cite{T1566}) (step \circled{1} in Figure \ref{fig:FIN6Steps}). After a host has been compromised, and command and control (C2) is established, FIN6 conducts internal reconnaissance, so as to find opportunities for privilege escalation, lateral movement and exfiltration in the second phase. This step relies on techniques for \emph{Account Discovery} (T1087 \cite{T1087}), \emph{Remote System Discovery} (T1018 \cite{T1018}), and \emph{System Network Configuration Discovery} (T1016 \cite{T1016}) (step \circled{2}). The next objective is Privilege Escalation: for this purpose, FIN6 leverages \emph{credential dumping} (T1003 \cite{T1003}) and \emph{access token manipulation} (T1134 \cite{T1134}) (step \circled{3}), and \emph{copies the Active Directory database file} (T1003.003 \cite{T1003.003}) in the Windows domain (step \circled{4}). The last step of this phase is the \emph{exfiltration of credentials in a compressed format} (T1560.001 \cite{T1560.001}), over \emph{SSH} as well as other channels (T1048 \cite{T1048}) (step \circled{5}).
In particular, FIN6 targets operational payment servers to \emph{inject malicious code} (T1055 \cite{T1055}) (step \circled{6}). Finally, the attack exfiltrates stolen data towards a remote server, using \emph{DNS tunneling} (T1048 \cite{T1048}) (step \circled{7}).

\vspace{3pt}
\noindent
\textbf{Attack emulation}. The second macro-phase of adversary emulation is the operational execution of the attack, leveraging the information provided by the previous phase. The attack techniques need to be associated with an implementation, provided by specific threat emulator tools. 
Among these emulators, MITRE CALDERA \cite{Applebaum2016IntelligentAR, CALDERAGitHub} is the most prominent one, as it is designed for orchestrating complex attacks that consist of multiple techniques. 
A threat emulator needs to be configured before running a simulation: we will refer to CALDERA to describe the necessary steps. First of all, it is necessary to setup the Command and Control server, which will coordinate the simulations. The next step is to deploy the remote agent on the target machine: an agent is a software program which connects to the C2 server via a predefined contact method \cite{CALDERAdocs}. The C2 server will be responsible for sending the instructions to the agents. Once the agent has been deployed, it is possible to choose the \textit{adversary profile} for the simulation. A profile is a group of abilities that includes the attack techniques typically used by a specific attacker \cite{CALDERAdocs}.

In the case of FIN6, a new adversary profile for CALDERA needs to be configured, by combining multiple techniques from CALDERA into an orchestrated attack. Therefore, the emulation of an attack relies on accurate planning of adversarial activities, which is currently performed by human operators. To overcome this issue, it is useful to automate the analysis of CTI, by extracting adversary techniques and mapping them to a standard taxonomy (e.g., MITRE ATT\&CK) to feed the emulation tool.




\section{Methodology}
\label{sec:Classification}

\begin{figure}[htbp]
\centering
\centerline{\includegraphics[width=\linewidth]{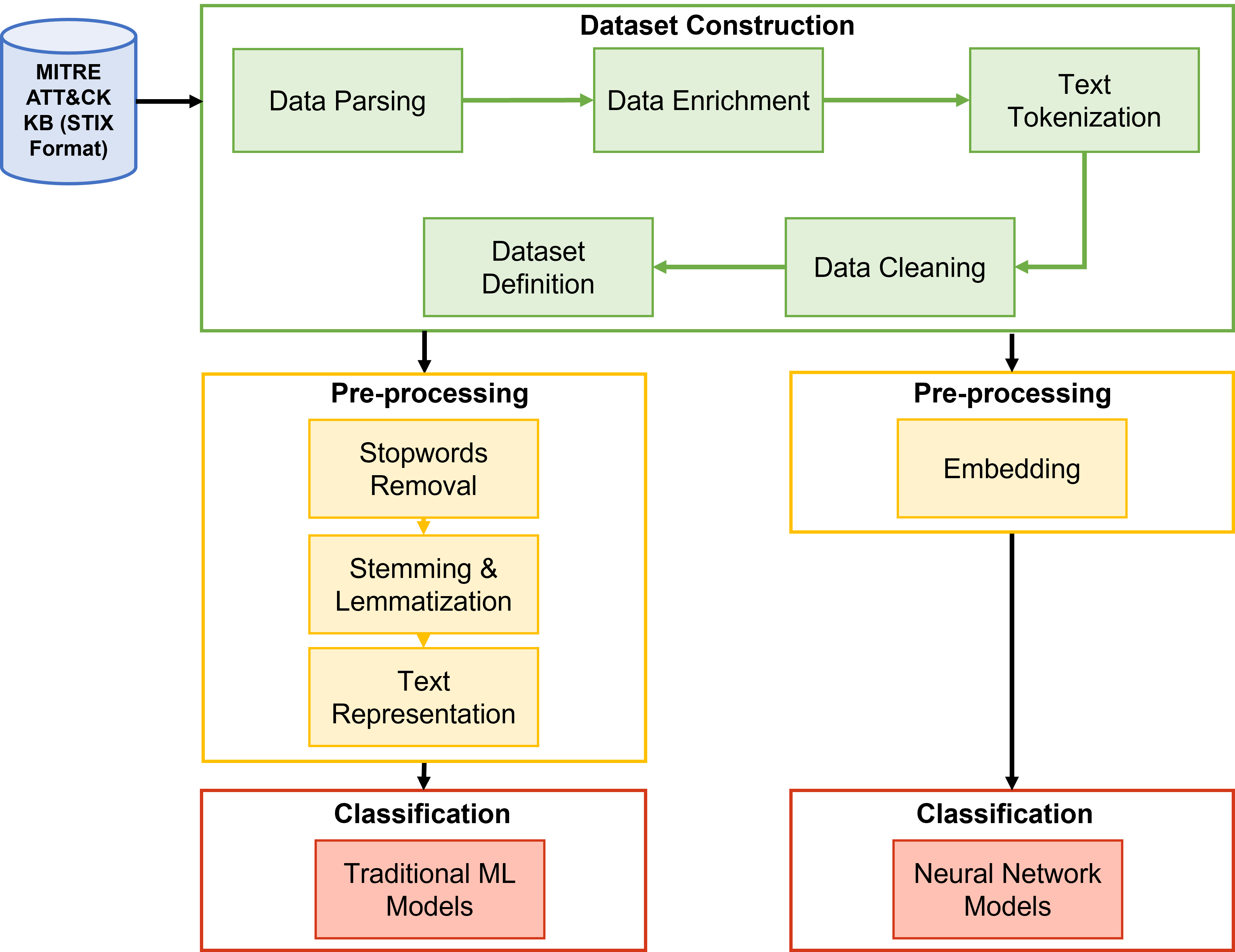}}
\caption{Experimental approach.}
\label{fig:Classification}
\end{figure}

In this section, we present the workflow of our experimental approach. Since our goal is to fit CTI into a finite set of categories (i.e., ATT\&CK techniques), we address this problem as a \emph{classification} machine learning task. Therefore, we build datasets to train and to evaluate machine learning algorithms, and perform several experiments on multiple classification techniques. 
Our experimental approach revolves around three phases: \textit{Dataset Construction}, \textit{Pre-processing}, and \textit{Classification}, shown in Figure \ref{fig:Classification}. The outcome will be a ML model ready to be used for the mapping of natural language documents into the corresponding techniques of the MITRE ATT\&CK framework.

In the \textit{Dataset Construction}, we built a new dataset, which consists of (unstructured) descriptions of adversarial techniques in natural language. Each sample in the dataset describes one malicious technique, and has been annotated with a label, i.e., a technique from the taxonomy of the MITRE ATT\&CK framework. 
We built the dataset using the public \emph{knowledge base} of the MITRE ATT\&CK framework. This knowledge base provides detailed descriptions in natural language of several threat actors and their malware campaigns. These descriptions are also mapped to MITRE ATT\&CK in a structured way. Therefore, we built the dataset by extracting both descriptions and their mappings to MITRE ATT\&CK. In particular, we analyzed the knowledge base as released by MITRE using the Structured Threat Information eXpression (STIX) language \cite{stix}.

STIX is a representation of CTI as serializable JSON. It describes cyber threats using an extensive set of elements, including: cybercriminal groups, their campaigns and targeted sectors and companies, their targeted vulnerabilities, their software tools, and any related artifacts, such as hash values of  malicious executables, IPs and domains involved in their campaigns, and other indicators of compromise (IOCs). These elements are represented as nodes (\textit{Domain Objects}) in a graph, and are connected by links (\textit{Relationship Objects}) to represent related concepts (e.g., a malware campaign is linked to its IOCs). These objects include attributes, such as a unique ID, a name and a description in natural language, and references to external documents with more information (e.g., online incident reports, and other attack taxonomies such as Common Attack Pattern Enumerations and Classifications (CAPEC) \cite{capec}). 

In the first step of our process, we \textit{parsed} the knowledge base in STIX format. In particular, we extracted information from the \textit{attack pattern} objects, which represent attack techniques from the MITRE ATT\&CK framework, and which include a general description of the technique in natural language. Moreover, we extracted the links (i.e., Relationship Objects) between the attack patterns and the threat actors and campaigns that adopted them (e.g., \textit{Intrusion Set} objects). These Relationship Objects include an additional description of the technique, in the context of a specific attack. Finally, in the case of attack pattern objects that provide external references to the CAPEC taxonomy, we used these references to obtain more descriptions from CAPEC (also available in STIX format) about the attack techniques.




The subsequent step in our approach was \textit{cleaning} the enriched data by removing all the unnecessary information, such as empty lines, and links and references to external sources. After cleaning the data, we proceeded with \textit{Text Tokenization}: the text was tokenized at sentence-level using the Natural Language Toolkit (NLTK) \cite{bird2009natural} sentence tokenizer. Finally, we tied all this information together to \textit{define} the dataset. Each entry contains the following data: ATT\&CK technique ID, name, and a description of the technique in natural language, as shown in Table \ref{tab:Dataset Examples}.

We release the dataset as open data\footnote{\url{https://github.com/dessertlab/cti-to-mitre-with-nlp}}. The extent of the resulting dataset is reported in Table \ref{tab:Dataset-Composition}. All 188 techniques in the MITRE ATT\&CK framework were covered. Each technique can appear in the dataset multiple times with different descriptions, as a technique may have been adopted in multiple attack campaigns, even if in different ways. Each sample describes a technique with one or multiple sentences. In total, 12,945 samples were included.

\begin{table}[h!]
\caption{Dataset of CTI samples in numbers.}
\begin{center}
\begin{tabular}{cc}
    \hline
    \textbf{Categories} & 188 \\
    \hline
    \textbf{Samples} & 12,945 \\
    \hline
    \textbf{Unique words} & 7,881 \\
    \hline
    \textbf{Total \# of words} & 193,453 \\
    \hline
\end{tabular}
\label{tab:Dataset-Composition}
\end{center}
\end{table}

\renewcommand{\arraystretch}{1.5}
\begin{table*}[htbp]
\caption{Example of dataset entries.}
\begin{center}
\resizebox{\linewidth}{!}{%
\begin{tabular}{p{0.35\textwidth}>
{\centering}p{0.15\textwidth}>
{\centering}p{0.2\textwidth}>
{\centering\arraybackslash}p{0.23\textwidth}}
    \hline
    \textbf{Text Description} & \textbf{Technique} & \textbf{Sub-technique} & \textbf{Sub-technique name}\\
    \hline
    Hydraq creates a backdoor through which remote attackers can adjust token privileges & T1134 & T1134 & Access Token Manipulation \\
    \hline
    XCSSET attempts to discover accounts from various locations such as a user's Evernote, AppleID, Telegram, Skype, and WeChat data & T1087 & T1087 & Account Discovery \\
    \hline
    PoisonIvy creates a Registry key in the Active Setup pointing to a malicious executable & T1547 & T1547.014 & Active Setup \\
    \hline
    BADNEWS can use multiple C2 channels, including RSS feeds, Github, forums, and blogs & T1102 & T1102.002 & Bidirectional Communication\\
    \hline
    Hildegard has searched for SSH keys, Docker credentials, and Kubernetes service tokens & T1552 & T1552.001 & Credentials In Files \\
    \hline
    Kobalos can write captured SSH connection credentials to a file under the /var/run directory with a .pid extension for exfiltration & T1074 & T1074 & Data Staged \\
    \hline
    TEARDROP was decoded using a custom rolling XOR algorithm to execute a customized Cobalt Strike payload & T1140 & T1140 & Deobfuscate/Decode Files or Information \\
    \hline
    It has also disabled Windows Defender's Real-Time Monitoring feature and attempted to disable endpoint protection services & T1562 & T1562.001 & Disable or Modify Tools \\
    \hline
    APT29 has exfiltrated collected data over a simple HTTPS request to a password-protected archive staged on a victim's OWA servers & T1048 & T1048.002 & Exfiltration Over Asymmetric Encrypted Non-C2 Protocol \\
    \hline
\end{tabular}}
\label{tab:Dataset Examples}
\end{center}
\end{table*}
Once the dataset has been built, it needs to be \textit{pre-processed}. The Pre-processing phase encompasses different steps depending on the type of the machine learning model.
For traditional machine learning (ML) models, this phase starts with \textit{Stopwords Removal}. The following step is \textit{Stemming and Lemmatization}: stemming consists in removing all the prefixes/suffixes from the words, while lemmatization links all the inflected words to the corresponding lemma (e.g., \textit{exploits} and \textit{exploiting} are both linked to the lemma \textit{exploit}). The last step of this phase concerns the \textit{Text Representation}: the information contained in a sentence is converted to a bag of words, with each sentence encoded as a one-hot vector. Using bags of words results in large feature vectors: this may cause loss of the word context and position inside the sentence. In a large corpus, it is common to have more frequent words, related to language patterns, that do not bring useful features to the classifier and less frequent yet informative words. To balance the representation, we can apply Term-Frequency Inverse Document-Frequency (TF–IDF) term weighting to transform count features into floating point values, preserving meaningful but rarer information present in the entire corpus.
On the contrary, for neural network models, the pre-processing phase consists of representing text with word \textit{embeddings}: this type of representation encodes each word taking the context into account. In this way, it is possible to define the similarity between two words, by means of their distance.

After this step, the data is ready for \textit{classification}. We consider several classification models, among two sets: 
\begin{itemize}
    \item \textit{Traditional ML models}. This set includes well-studied classifiers used for NLP tasks since decades, including: Naive Bayes, Logistic Regression, Support Vector Machines (SVM), and Multi-Layer Perceptron (MLP); 
    \item \textit{Deep Neural Networks}. Following the latest developments in NLP research, we consider recent classifiers based on complex neural network architectures, including: Recurrent Neural Networks (RNN), in particular Long Short-Term Memory (LSTM), Convolutional Neural Networks (CNN), and Transformers \cite{vaswani2017attention}.
\end{itemize}

In particular, among the deep learning-based classifiers, the \emph{Transformer architecture} represents the most recent advancement in the field of NLP research. It introduces the \textit{self-attention} mechanism, which weights the tokens of the input data according to their importance. Therefore, we also build a classifier using \emph{SecureBERT}, a Language Model (LM) pre-trained on cybersecurity terms \cite{secBert}. It is based on BERT, a multi-layer bidirectional Transformer encoder, composed of 12 layers (transformer blocks), 12 attention heads, and 110 million parameters in its \textit{base} version. Generally pre-trained on Masked Language Modeling (MLM) and Next Sentence Prediction (NSP) tasks, SecureBERT presents only a MLM head layer.

Our attack classification problem poses some challenges to ML classifiers. Since we deal with a large number of different classes (i.e., the 188 MITRE ATT\&CK techniques), 
we are addressing a multi-class classification problem with a quite large number of categories. It is worth mentioning that text classification problems are usually binary, or have a few categories (typically around 10). 
Moreover, we also observed that the attack data is quite imbalanced: the majority class is the \textit{Command and Scripting Interpreter} (T1059) technique, which is represented by 698 occurrences, while the minority class is \textit{Cloud Service Dashboard} (T1538), which occurs only 4 times. The different frequencies of the techniques reflect their actual adoption by threat actors, and their time of introduction in the ATT\&CK Framework, as recent techniques tend to be less represented. 
These challenges motivate our experimental study on ML classifiers for the problem of mapping CTI, in order to identify their relative strengths and limitations.

\section{Experimental evaluation with cross-validation}
\label{sec:Results}

In this section, we perform a preliminary evaluation on the dataset itself, by adopting a cross-validation approach. In the following, we elaborate on the evaluation methods and experimental results.


\subsection{Evaluation metrics and baseline}
\label{sec:Metrics}
In order to compare the performance of the classification models, we considered four different metrics. Before introducing metrics, we briefly recap some key concepts: \textit{True Positives} (TP) and \textit{True Negatives} (TN) represent a correct prediction made by the classifier, respectively for positive and negative classes, thus the predicted label matches the actual one. On the other hand, \textit{False Positives} (FP) and \textit{False Negatives} (FN) respectively represent an entry from a positive/negative class that has been incorrectly classified. Based on these definitions, we analyze the following metrics for multi-class classification \cite{Grandini2020MetricsFM}: \textit{precision} is the number of correct positive predictions (TP) divided by the total number of positive predictions (TP + FP) made by the model, while \textit{recall} is the number of true positives (TP) divided by the total number of positive instances (TP + FN). Precision indicates how much the model can be trusted when it predicts an instance as positive, and recall shows the ability of the model to find all the positive instances in the dataset.

\begin{displaymath}
    precision = \frac{TP}{TP + FP}
\end{displaymath}
\begin{displaymath}
    recall = \frac{TP}{TP + FN}
\end{displaymath}


Given the imbalanced nature of our dataset and the skewed distribution of classes, we considered the \textit{F-Measure} statistic instead of the accuracy. Also known as \textit{F1-Score}, The F-Measure is the armonic mean of precision and recall, and tells us how well the model is able to classify all classes, including minority ones. 

\begin{displaymath}
    F\mhyphen Measure = 2\cdot\frac{precision \cdot recall}{precision + recall}
\end{displaymath}

In some use cases, the ML models could be used as a \emph{recommender}, to suggest to the analyst a set of candidate attack mappings (instead of providing only one ``hard'' prediction). Therefore, as the last metric, we analyzed the \textit{top K accuracy} (AC@K), with K equal to 3: it indicates how likely the correct class is ranked in the top 3 predictions. 
We considered the weighted version of these metrics, so as to take into account the frequency of each class.

As baseline for our evaluation, we consider the Threat Report ATT\&CK Mapping project (TRAM) \cite{TRAMGitHub}, an open source platform developed by the CTID. This platform maps ATT\&CK techniques from unstructured text, similarly to our work. At the time of writing, the TRAM project only supports three classifiers (Multinomial Naive Bayes, a basic Logistic Regression model, and MLP), and no systematic experimental evaluation has yet been reported in the scientific literature. Moreover, the TRAM project only provides a limited dataset of CTI-to-ATT\&CK mappings, which only covers $80$ attack techniques with more than 1 sample, with a total of $1,482$ samples of attack descriptions.
In this work, we consider a wider set of ML algorithms, and we introduce a new, larger dataset for the evaluation.




\subsection{Model Training}
\label{sec:Model Training}

We developed the processing pipeline using the SKLearn Library \cite{scikit-learn,sklearn_api}, which provides several resources for ML. The hyper-parameters of the models were tuned according to the imbalance of our dataset. In a preliminary data analysis, we noticed that the vocabulary included about 7,800 unique words, resulting in very large features vectors. 
As explained in Section \ref{sec:Classification}, we applied traditional NLP pre-processing steps: stemming, lemmatization and stopwords removal. Then, we used TFIDFVectorizer, based on TF-IDF statistic, to obtain a numerical representation of the text, reflecting the importance of each word in the corpus. Besides, we set the \textit{ngram\_range} to $(1,2)$, to extract both unigrams and bigrams, and \textit{max\_features} to $10,000$.

For the Naive Bayes and Logistic Regression models, we presented two different versions: the LogisticRegression with the \textit{class\_weight} parameter set to \textit{balanced} and the ComplementNaiveBayes, which take into account the imbalanced representation of classes, and the basic library version of LogisticRegression and MultinomialNaiveBayes.
For SVM, we consider both the case of One versus One (OvO), which trains $n(n-1)/2$ binary classifiers to split a multi-class classification into multiple binary classification problems; and One versus the Rest (OvR), which instantiates \textit{n} classifiers to classify a sample between a single category versus all of the other choices united. 
We chose a linear kernel and set the \textit{class\_weight} parameter to \textit{balanced} for both of them.
Finally, for MLP, we used the model provided by SKLearn, by setting up the maximum number of iterations for the learning algorithm and the early stopping parameter to prevent overfitting.

For the deep learning models, we leveraged the Keras library \cite{chollet2015keras}. We focused on both RNN and CNN models, using word embeddings as text representation. 
In these models, the first layer transforms words into their equivalent embedding vectors: each sentence has been previously split into tokens and subsequently padded or truncated to meet a maximum length constraint. 
We chose to implement the LSTM architecture, a RNN model commonly used for text processing tasks. The hyper-parameters setup for the LSTM layer were: 150 for the number of units, defining the output dimension, and 0.2 for the dropout parameter, useful to prevent overfitting. Furthermore, we added a slightly different version of this model that initializes the embedding layer with a pre-trained cybersecurity-specific Word2vec model \cite{padia-etal-2018-umbc}. This model provides a word representation with a hundred size vectors: we extended its vocabulary and retrained the model with an incremental approach to add new words.

On the other hand, for the CNN, we set up a single layer composed of 256 filters to extract features and a convolution window of 5, followed by a Global Max Pooling layer. 
Both models present a final fully connected layer of 188 units for the classification task. To avoid overfitting, we took advantage of Keras \textit{EarlyStopping} callback to automatically define the number of epochs during the training of the models. The batch size was set to 64, while the validation-set percentage to 0.1. 

Finally, we built a classifier based on Transformers, by means of the transfer learning technique and the pre-trained SecureBERT model \cite{secBert} \cite{secBert_model}. Leveraging the Hugging Face \cite{huggingface} Framework and the Pytorch \cite{pytorch} library, we fine-tuned the model on our dataset and re-trained it with a very low learning rate, $1e-05$, so as to adapt pre-trained features to our data. SecureBERT presents a BERT core followed by a MLM head layer. We extracted the pre-trained BERT core and built a classifier on top of that adding two layers: a dropout layer, with a fraction of $0.3$, to prevent overfitting, and linear layer for the classification into 188 different categories. We set the batch size to $16$ and the number of epochs to $10$.

To train each model, the data set was split into two parts: 80\% for the training set and 20\% for the test set. We used a stratified split to preserve the same proportion of examples in both the training and test sets.

\subsection{Results}
\label{subsec: class_results}
We first consider the results obtained from our dataset, presented in Table \ref{tab:Our Dataset Results}.
The best result is achieved by SecureBERT, according to both F-Measure and AC@3, followed respectively by MLP and the OVR version of the SVM model. For both Naive Bayes and Logistic Regression models we presented the results of multiple variants, as discussed in Section \ref{sec:Model Training}. The aim of this choice was to demonstrate how the models showed better results when taking the frequency of each class into account. 
As far as SVM models are concerned, we notice that there is no significant difference between the performance of the OvO and OvR strategies. This result supports the choice of the OvR strategy instead of the OvO one, since it constructs a linear number of classifiers rather than a quadratic one. SVM show good results both in the case of F-Measure and top K accuracy. These values are very close to the top ranking model for both these metrics. 
The other DL-based classifiers did not match the traditional ML algorithms' performance, as we observed a 10\% gap among them. It is worth noting that the LSTM model performed slightly better when the embedding layer was initialized with the pre-trained cybersecurity Word2vec weights. 

\newcolumntype{X}{>{\centering\arraybackslash}m{2cm}} 
\begin{table}[htbp]
\caption{Evaluation of the results in our dataset.}
\begin{center}
\begin{tabular}{Xcccc}
    \hline
    \multirow{2}{*}{\textbf{Models}}&\multicolumn{4}{c}{\textbf{Metrics}} \\
    \textbf{} & \textbf{F-Measure}& \textbf{Precision} & \textbf{Recall} & \textbf{AC@3} \\
    \hline
    Complement Naive Bayes & 63.9\% &  65.9\% & 66.3\% & 66.6\% \\
    \hline
    Multinomial Naive Bayes & 36.8\% & 49.2\% & 41.5\% & 20.2\% \\
    \hline
    Logistic Regression & 55.6\% & 59.1\% & 60.5\% & 43.6\% \\
    \hline
    Logistic Regression (balanced) & 64.6\% & 69.3\% & 64.5\% & 79.6\% \\
    \hline
    SVM (OvO) & 69.9\% &  71.8\% & 70.2\% & 77.2\% \\
    \hline
    SVM (OvR) & 69.9\% & 71.8\% & 70.2\% & 77.3\% \\
    \hline
    MLP & 70.4\% & 71.9\% & 71.6\% & 77.3\% \\
    \hline
    LSTM & 57.7\% & 59.1\% & 58.5 \% & 54.7\% \\
    \hline
    LSTM (Word2vec) & 61.0\% & 62.2\% & 62.3\%& 64.4\% \\
    \hline
    CNN & 61.4\% & 63.0\% & 62.7\% & 59.5\% \\
    \hline
    SecureBERT & \textbf{72.5\%} & 72.5\% & 72.5\% & \textbf{86.9\%} \\
    \hline
\end{tabular}
\label{tab:Our Dataset Results}
\end{center}
\end{table}

\begin{table}[htbp]
\caption{Evaluation of the results in the TRAM dataset.}
\begin{center}
\begin{tabular}{Xcccc}
    \hline
    \multirow{2}{*}{\textbf{Models}}&\multicolumn{4}{c}{\textbf{Metrics}}\\
    \textbf{} & \textbf{F-Measure}& \textbf{Precision} & \textbf{Recall} & \textbf{AC@3} \\
    \hline
    Complement Naive Bayes & 53.7\% &  59.1\% & 61.6\% & 57.0\% \\
    \hline
    Multinomial Naive Bayes & 23.5\% & 29.6\% & 31.9\% & 16.7\% \\
    \hline
    Logistic Regression & 40.9\% & 50.6\% & 46.4\% & 31.0\% \\
    \hline
    Logistic Regression (balanced) & 57.7\% & 66.8\% & 58.9\% & 65.2\% \\
    \hline
    SVM (OvO) & 60.9\% &  63.8\% & 64.6\% & 44.0\% \\
    \hline
    SVM (OvR) & \textbf{60.9\%} & 63.8\% & 64.6\% & 42.6\% \\
    \hline
    MLP & 50.5\% & 56.0\% & 54.2\% & 42.0\% \\
    \hline
    LSTM & 34.3\% & 36.0\% & 36.3\% & 32.4\% \\
    \hline
    LSTM (Word2vec) & 36.5\%& 37.2\%& 39.7\%& 40.2\% \\
    \hline
    CNN & 42.4\% & 42.1\% & 47.1\% & 38.5\% \\
    \hline
    SecureBERT & 52.5\% & 52.5\% & 52.5\% & \textbf{68.6\%} \\
    \hline
\end{tabular}
\label{tab:TRAM Dataset Results}
\end{center}
\end{table}

\begin{table}[htbp]
\caption{Evaluation of results in mixed dataset, ours combined with TRAM.}
\begin{center}
\begin{tabular}{Xcccc}
    \hline
    \multirow{2}{*}{\textbf{Models}}&\multicolumn{4}{c}{\textbf{Metrics}}\\
    \textbf{} & \textbf{F-Measure}& \textbf{Precision} & \textbf{Recall} & \textbf{AC@3} \\
    \hline
    Complement Naive Bayes & 62.3\% &  64.0\% & 64.8\% & 63.2\% \\
    \hline
    Multinomial Naive Bayes & 36.5\% & 50.5\% & 41.3\% & 36.5\% \\
    \hline
    Logistic Regression & 56.8\% & 60.0\% & 61.2\% & 42.2\% \\
    \hline
    Logistic Regression (balanced) & 64.2\% & 68.6\% & 63.9\% & 79.6\% \\
    \hline
    SVM (OvO) & 68.5\% &  70.3\% & 69.0\% & 73.1\% \\
    \hline
    SVM (OvR) & 68.5\% & 70.3\% & 69.0\% & 73.0\% \\
    \hline
    MLP & 68.9\% & 69.8\% & 70.0\% & 69.9\% \\
    \hline
    LSTM & 57.3\% & 58.6\% & 58.2\% & 52.9\% \\
    \hline
    LSTM (Word2vec) & 59.8\%& 60.4\%& 61.0\%& 59.8\% \\
    \hline
    CNN & 62.4\% & 62.7\% & 64.1\% & 58.0\% \\
    \hline
    SecureBERT & \textbf{71.6\%} & 71.6\% & 71.6\% & \textbf{86.1\%} \\
    \hline
\end{tabular}
\label{tab:Mixed Dataset Results}
\end{center}
\end{table}

We then evaluated the ML algorithms using the dataset from the TRAM project. The comparison with the TRAM project serves as a baseline for the evaluation, and provides evidence of the industry's interest in the problem addressed by this work. 
The TRAM dataset comes with a smaller absolute number of samples compared to our proposed dataset. As we can observe from the metrics in Table \ref{tab:TRAM Dataset Results}, the poorest performance is achieved by the DL-based models, which adopt word embeddings to represent text, and which benefit from large training datasets. These models are penalized when evaluated on the smaller TRAM dataset, which contains a lower number of sample sentences compared to our dataset. Similarly, while maintaining the best ranking for AC@3, SecureBERT is outperformed by the SVM classifier in terms of F-Measure. 
Traditional models, which use BoW, show better results: the SVM model is the best model according to the F-Measure, while Logistic Regression is the one with the best trade-off between F-Measure and AC@3.

To complete our evaluation, we combined the TRAM dataset with ours. 
Table \ref{tab:Mixed Dataset Results} shows the results of the classification for the extended dataset. SecureBERT, as for the evaluation on our dataset, remains the best model, providing the best score in both F-Measure and AC@3. The majority of the traditional models continue to provide good results: the balanced Logistic Regression model outperforms all the other ones scoring as the second best for the AC@3, while MLP has the second higher F-Measure. The performance is comparable with the first analysis, with a small decrease of the metrics as an artifact of the cross-validation process, and due to a higher variety of data in the combined dataset. 

Although the performance of the classifier based on SecureBERT is surely promising and provides good results, it is worth mentioning that this ML model, based on the Transformer architecture, needs a longer training process, leveraging GPU power to handle a greater number of parameters.
In the end, the resulting model is far more complex in terms of memory occupancy. 



\subsection{Qualitative Error Analysis}
\label{sec:Analysis}


\renewcommand{\arraystretch}{1.5}
\begin{table*}[h!t]
\caption{Examples of mispredictions.}
\begin{center}
\resizebox{\linewidth}{!}{%
\begin{tabular}{p{0.35\textwidth}>
{\centering}p{0.1\textwidth}>
{\centering}p{0.1\textwidth}>
{\centering}p{0.2\textwidth}>
{\centering\arraybackslash}p{0.2\textwidth}}
    \hline
    \textbf{Text to Evaluate} & \textbf{Actual Technique} & \textbf{Predicted Technique} & \textbf{Actual Name} & \textbf{Predicted Name}\\
    \hline
    SharpStage has the ability to create persistence for the malware using the Registry autorun key and startup folder. & T1547 & T1112 & Boot or Logon Autostart Execution & Modify Registry\\
    \hline
    Stuxnet can create registry keys to load driver files. & T1112 & T1547 & Modify Registry & Boot or Logon Autostart Execution\\
    \hline
    A Gamaredon Group file stealer can transfer collected files to a hardcoded C2 server. & T1041 & T1020 & Exfiltration Over C2 Channel & Automated Exfiltration\\
    \hline
    However, an adversary may hijack the syscall interface code stubs mapped into a process from the vdso shared object to execute syscalls to open and map a malicious shared object. & T1055 & T1569 & Process Injection & System Services\\
    \hline
    FIN6 has used has used Metasploit’s named-pipe impersonation technique to escalate privileges. & T1134 & T1036 & Access Token Manipulation & Masquerading\\
    \hline
    APT29 added their own devices as allowed IDs for active sync using Set-CASMailbox, allowing it to obtain copies of victim mailboxes. & T1098 & T1210 & Account Manipulation & Exploitation of Remote Services\\
    \hline
    Sowbug identified and extracted all Word documents on a server by using a command containing *.doc and *.docx. & T1083 & T1119 & File and Directory Discovery & Automated Collection\\
    \hline
\end{tabular}}
\label{tab:mispredictions}
\end{center}
\end{table*}

Considering the promising results obtained from the previous analysis, we qualitatively analyzed mispredictions, to evaluate how far they were from the choices of a human analyst. 
Following the intuition that the DL-based classifiers could be the ones worth deepening for further improvements, we collected all of CNN predictions on the test set. We manually analyzed several cases of misprediction: a representative subset is reported in Table \ref{tab:mispredictions}. 
In the first sample, the actual technique refers to an attacker's intervention in the system configuration to start a certain malicious program at startup. The predicted class is about the opponent's intervention on the Windows registry to hide a malicious configuration. The predicted action is implied by the correct one, because autostart execution for persistence is achieved by means of a Registry Key modification. In the second sample in the table, the classifier was mistaken in reverse, making a wrong prediction due to the Registry Key being mentioned in the description. 
We also observed that about the 46\% of mispredictions belong to the same tactic of the correct technique, such as in the third sample in Table \ref{tab:mispredictions}, where both labels are related to the \emph{Exfiltration} tactic of MITRE ATT\&CK. In these cases, the ML model was still able to extract relevant information about the goal of an attack technique. 
In light of the difficulty of our classification task (i.e., identifying an attack technique among a very large set of techniques), this is a relevant result. 
We also found that several sentences are ambiguously worded and could be correctly mapped to multiple ATT\&CK techniques, even by a human analyst. The fourth, fifth, and sixth example in Table \ref{tab:mispredictions} could potentially be associated with both the predicted and the actual label. In the last example, the description seems more likely related to the predicted label rather than the actual one. 
In conclusion, we saw how difficult is to precisely evaluate the effectiveness of the prediction made by the classifier based on a single phrase. From here, the necessity to conduct a more thorough evaluation, based on real-world CTI documents.

\section{Experimental evaluation on CTI documents}
\label{sec:Document}
\newcolumntype{X}{>{\centering\arraybackslash}m{2cm}} 
\begin{table*}[htbp]
\caption{Dataset of whole documents with Cyber Threat Intelligence.}
\begin{center}
\small
\resizebox{\linewidth}{!}{%
\begin{tabular}{p{0.35\textwidth}>
{\centering}p{0.15\textwidth}>
{\centering}p{0.2\textwidth}>
{\centering\arraybackslash}p{0.23\textwidth}}
    \hline
    \textbf{Threat} & \textbf{Reference} & \textbf{\# Techniques} & \textbf{\# Sentences}\\
    \hline
    Follow The Money: Dissecting the Operations of the Cyber Crime Group FIN6 & \cite{FIN6FireEye} & 17 & 81 \\
    \hline
    Pick-Six-Intercepting a FIN6 Intrusion, an Actor Recently Tied to Ryuk and LockerGoga Ransomware & \cite{FIN6Mandiant} & 50 & 101 \\
    \hline
    MenuPass: Japan-Linked Organizations Targeted in Long-Running and Sophisticated Attack Campaign & \cite{MenuPassSymantec} & 32 & 73 \\
    \hline
    MenuPass: United States v. Zhu Hua Indictment & \cite{MenuPassUS} & 23 & 46 \\
    \hline 
    WizardSpider: Ransomware Activity Targeting the Healthcare and Public Health Sector & \cite{WizardSpiderCISA} & 99 & 275 \\ 
    \hline 
    WizardSpider: Ryuk’s Return & \cite{WizardSpiderDFIR} & 72 & 76 \\ 
    \hline
\end{tabular}}
\label{tab:DatasetDoc Numbers}
\end{center}
\end{table*}

In the previous evaluation, we used ML models to classify short descriptions of the ATT\&CK techniques in natural language. The sample were short sentences about a technique engaged by an attacker. Nevertheless, in a real-world scenario, a human cybersecurity analysts deals with entire CTI documents, instead of individual sentences and techniques. In this case, the description in natural language spans all of the multiple techniques adopted in an attack campaign. 
For these reasons, we performed a document-level analysis, in which we would like to observe how many ATT\&CK techniques the classifiers are able to identify from a document. In this case, the output of the ML model is a set of techniques. In the previous case, every sample described exactly one ATT\&CK technique. We do not focus on mapping individual sentences to techniques; instead, we evaluate how many of the techniques presented in the document were correctly identified by the model. 

We built an additional new dataset for this evaluation. For each document describing a threat actor or attack, we identified a list of ATT\&CK techniques engaged in it. 
The documents were collected from the knowledge base of the MITRE ATT\&CK framework. ATT\&CK documents notorious malware campaigns and threat groups, by structuring information from incident reports and from cybersecurity experts. 
These sources typically provide a broad description of real attacks, including a general introduction, a description of the targeted industry sector, the general methodology of the threat actors, the technical information about attack techniques, and the suggested mitigations against the threat. 

\begin{figure*}[htbp]
     \centering
     \begin{subfigure}[b]{0.3\textwidth}
         \centering
         \includegraphics[width=\textwidth]{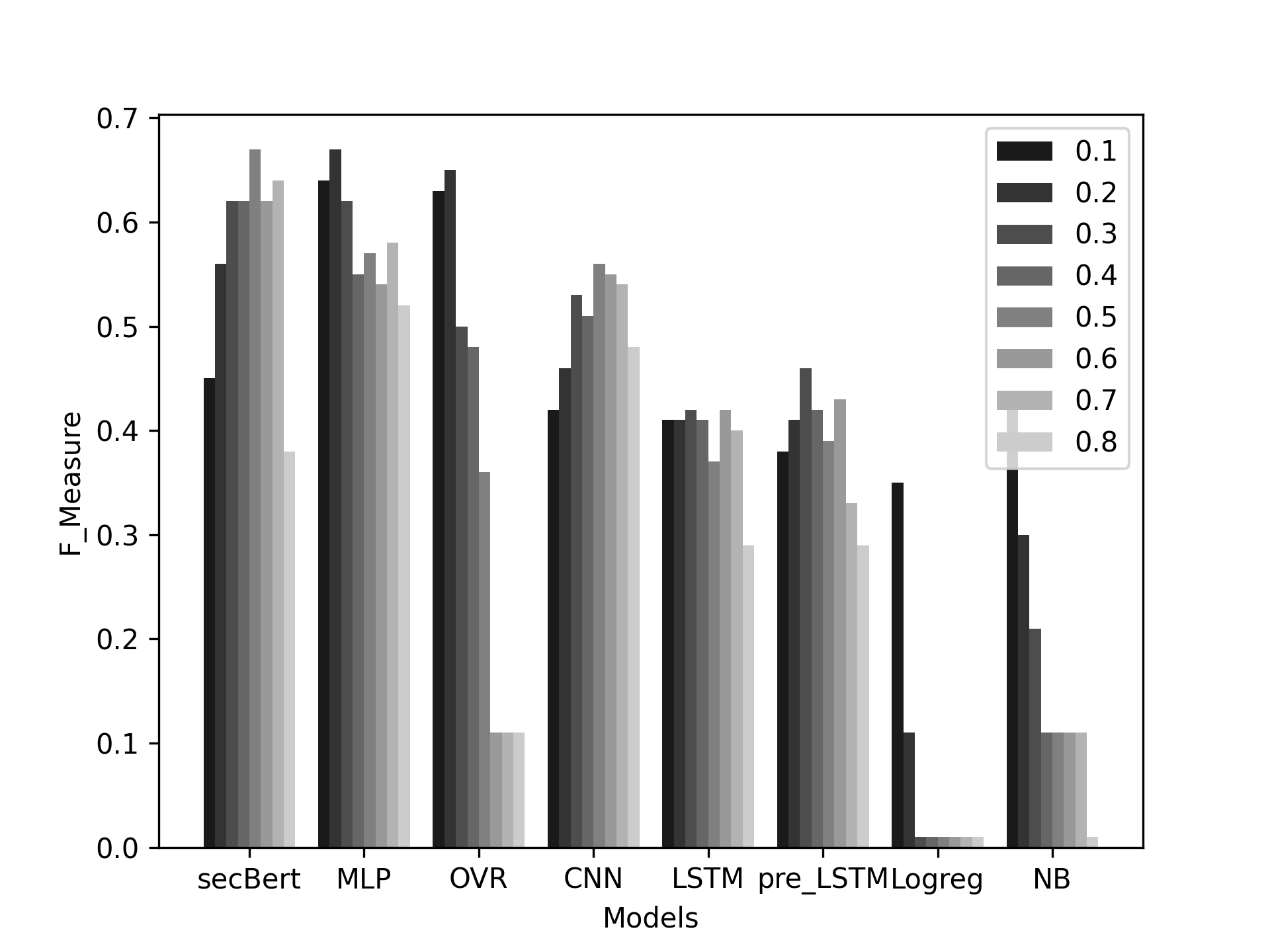}
         \caption{FIN6\cite{FIN6FireEye}}
         \label{fig:FIN6_ref1}
     \end{subfigure}
     \hfill
     \begin{subfigure}[b]{0.3\textwidth}
         \centering
         \includegraphics[width=\textwidth]{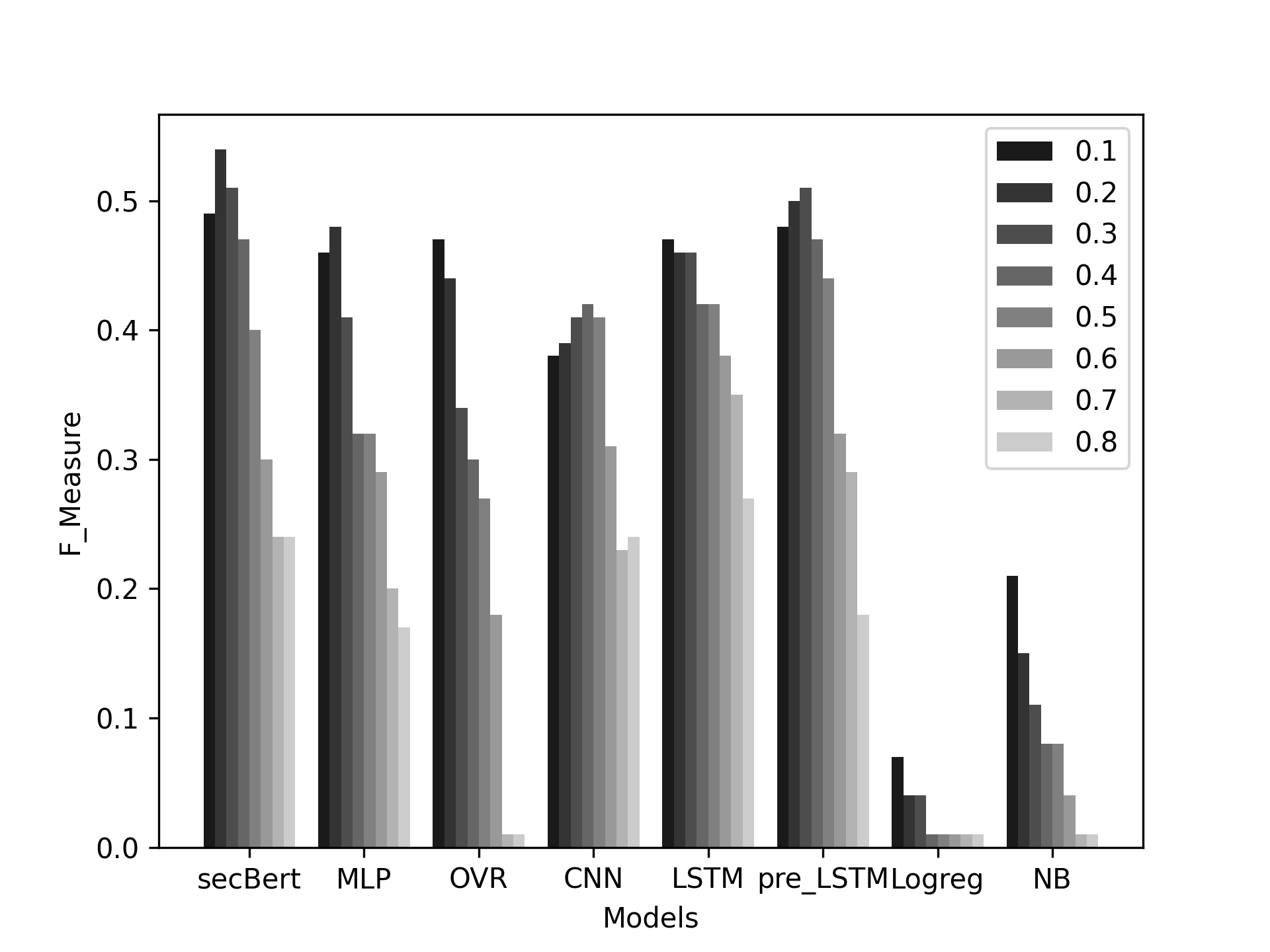}
         \caption{FIN6\cite{FIN6Mandiant}}
         \label{fig:FIN6_ref2}
     \end{subfigure}
     \hfill
     \begin{subfigure}[b]{0.3\textwidth}
         \centering
         \includegraphics[width=\textwidth]{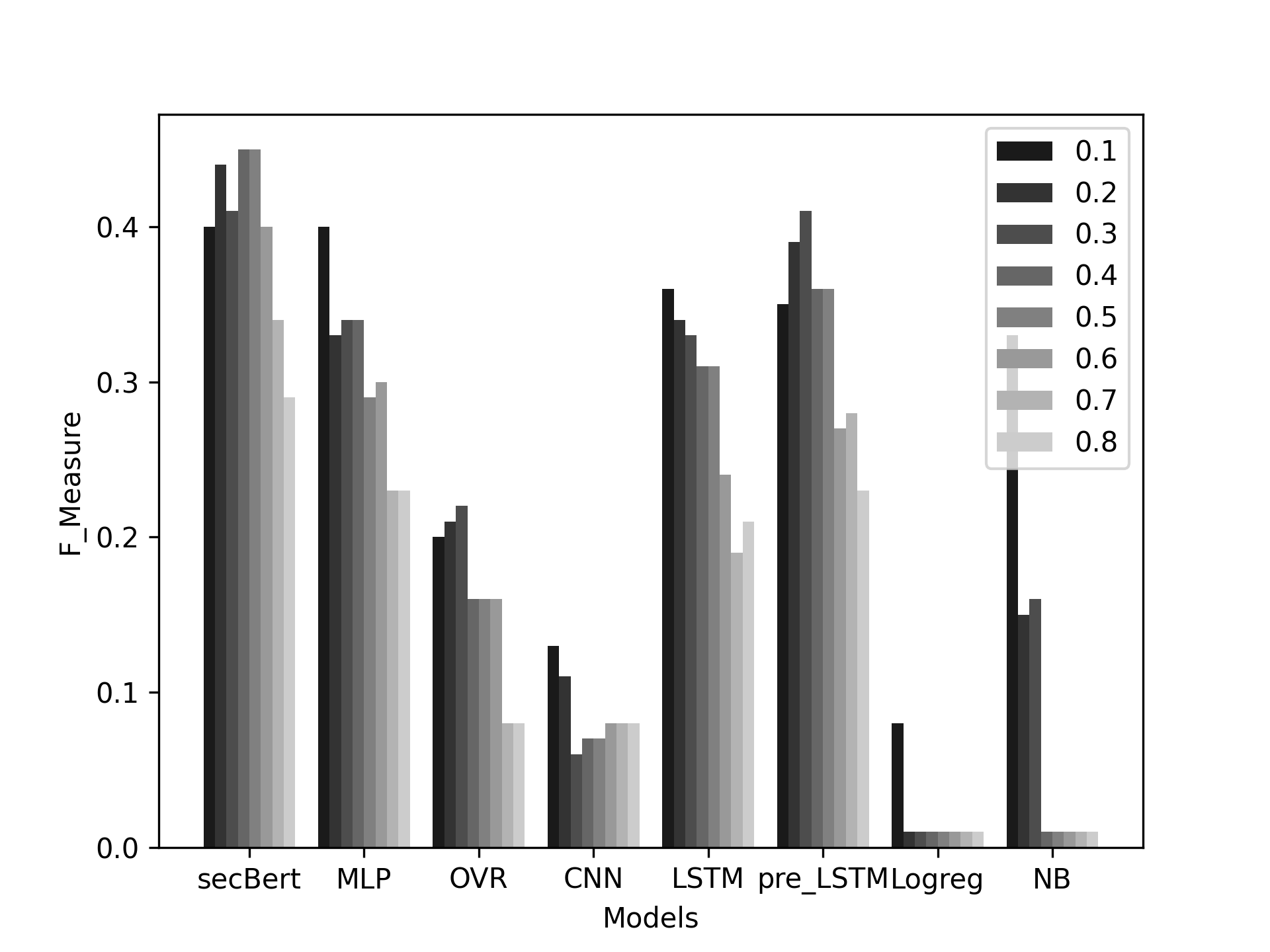}
         \caption{MenuPass\cite{MenuPassUS}}
         \label{fig:MenuPass_ref1}
     \end{subfigure}
     \begin{subfigure}[b]{0.3\textwidth}
         \centering
         \includegraphics[width=\textwidth]{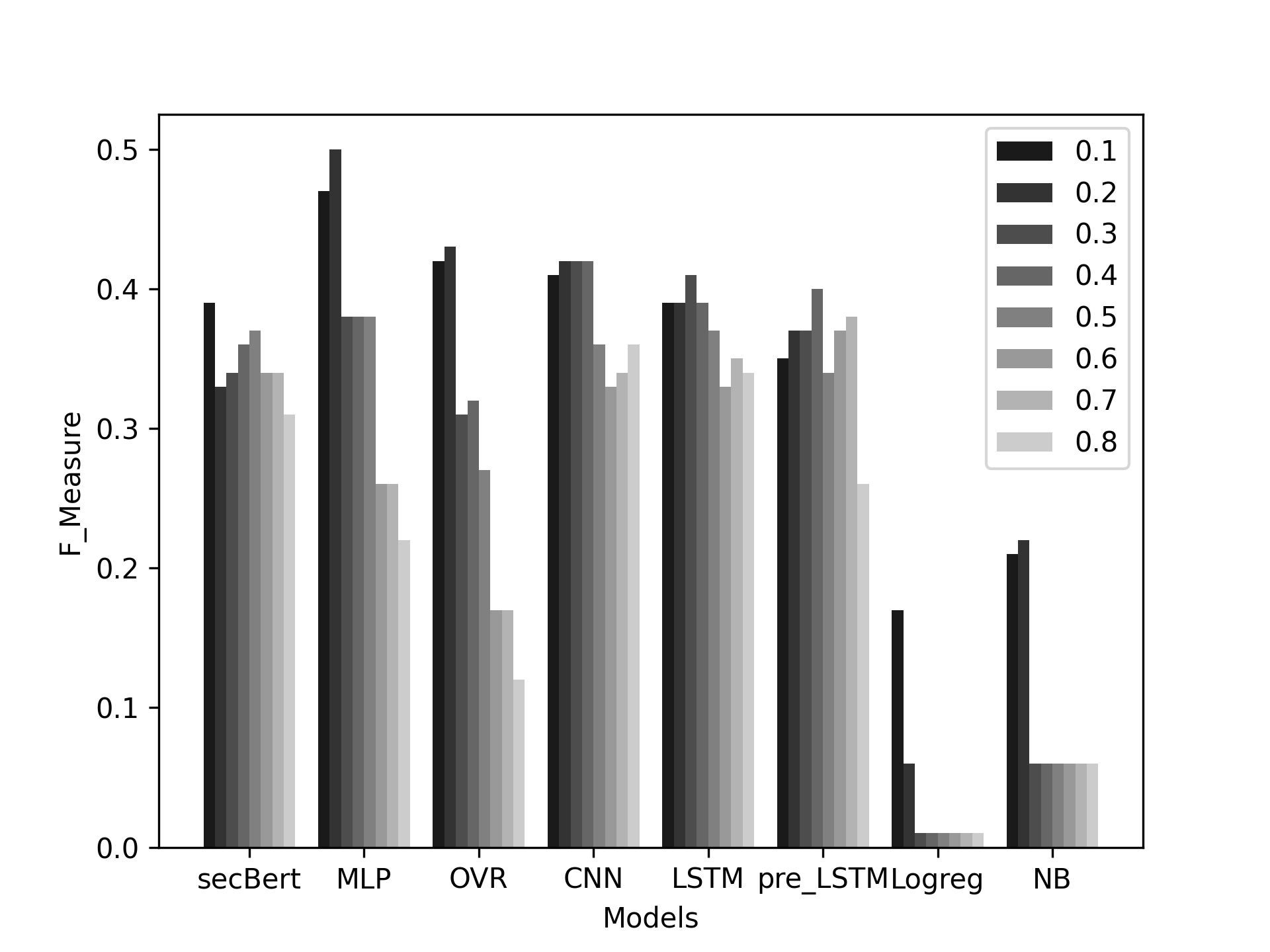}
         \caption{MenuPass\cite{MenuPassSymantec}}
         \label{fig:MenuPass_ref8}
     \end{subfigure}
     \hfill
     \begin{subfigure}[b]{0.3\textwidth}
         \centering
         \includegraphics[width=\textwidth]{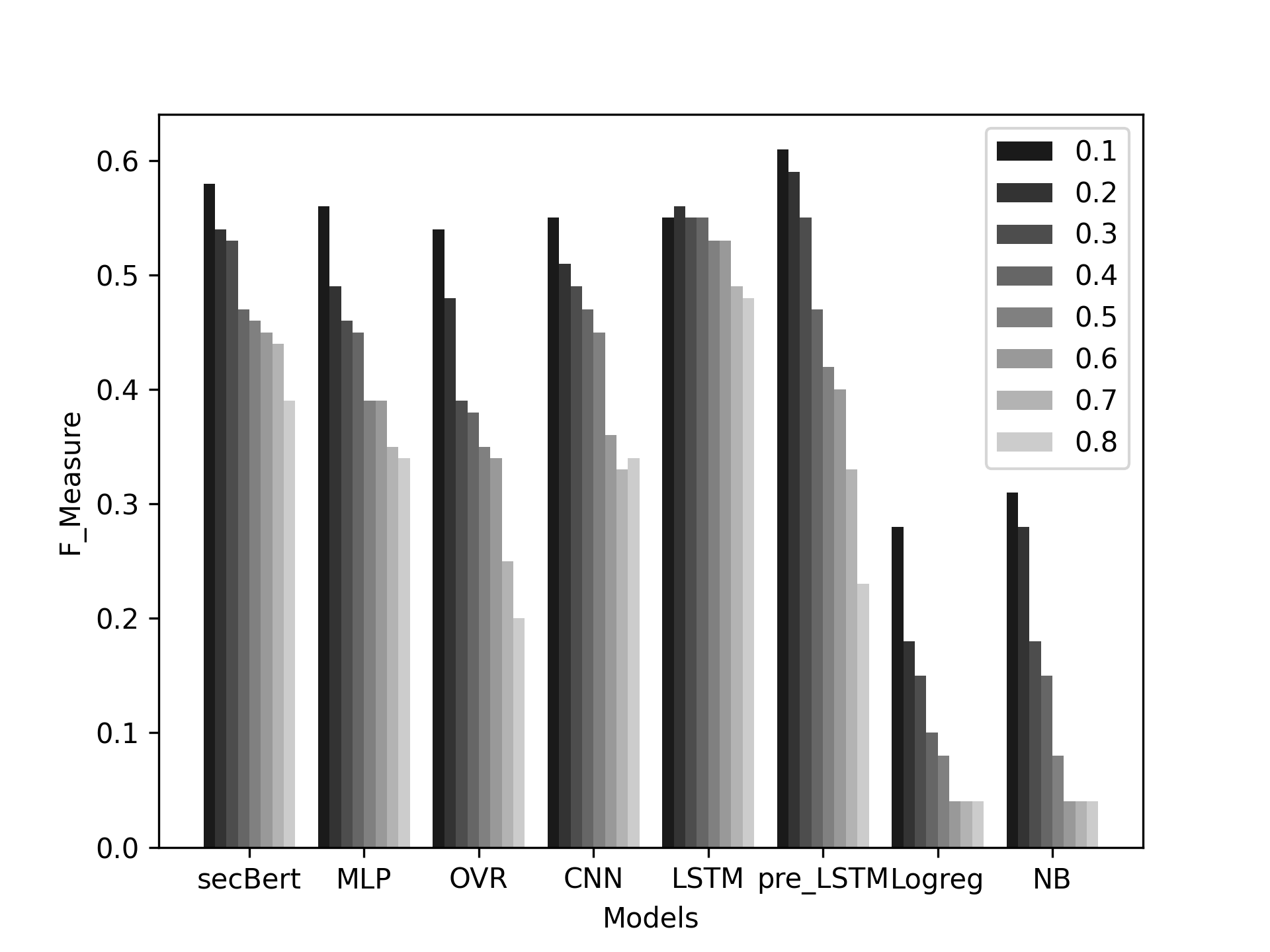}
         \caption{WizardSpider\cite{WizardSpiderCISA}}
         \label{fig:WizardSpider_ref2}
     \end{subfigure}
     \hfill
     \begin{subfigure}[b]{0.3\textwidth}
         \centering
         \includegraphics[width=\textwidth]{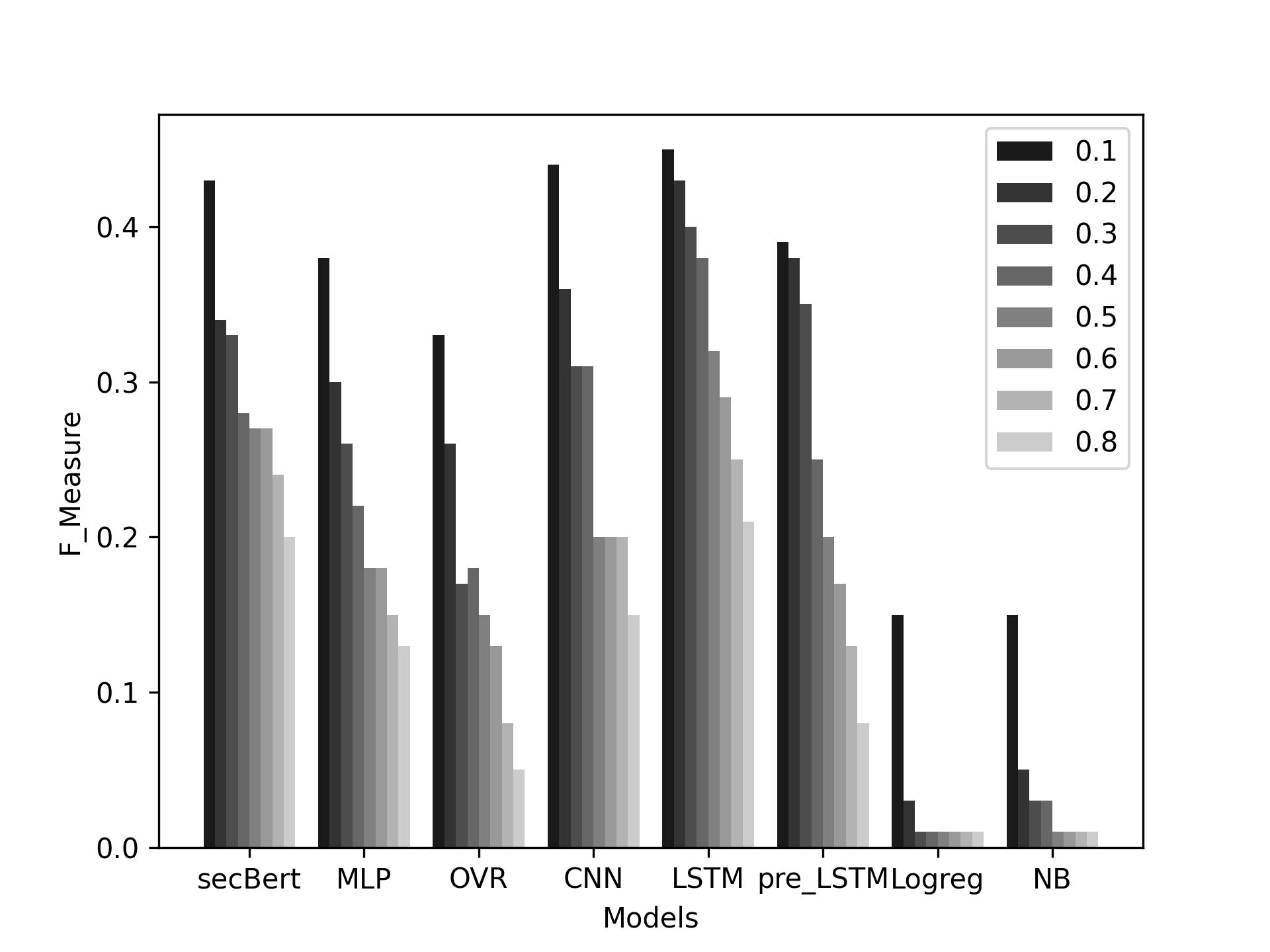}
         \caption{WizardSpider\cite{WizardSpiderDFIR}}
         \label{fig:WizardSpider_ref7}
     \end{subfigure}
        \caption{Document-level evaluation}
        \label{fig:Document-eval}
\end{figure*}

Differing from the first dataset (in which we looked at descriptions of individual techniques), in this case we followed references from the MITRE ATT\&CK to external information sources, and collected these sources as documents. Detailed information is provided in Table \ref{tab:DatasetDoc Numbers}: threat-related articles/reports, number of techniques described in them, and total number of sentences within each source. From the MITRE ATT\&CK knowledge base, we also derived for each document a ``ground-truth'' vector of techniques presented in the document. Then, we evaluated how many techniques our models are able to correctly predict, and the percentage of correct prediction overall.

To leverage ML models to identify multiple techniques from a document, we applied a threshold on the outputs of the classifier, i.e., a vector of probabilities, with one probability value for each ATT\&CK technique. If the probability exceeds the threshold, we include the corresponding technique in the set of ATT\&CK techniques chosen for the document. This approach also allows the ML model to distinguish between pertinent and non-pertinent sentences in the document (i.e., sentences not related to any technique), since a document typically includes sentences that are not strictly related to attack patterns. 
Finally, we use the predictions to compute metrics (precision, recall, and F-measure) in order to compare the results of different ML models. For this type of analysis, we defined \textit{precision} as:
\begin{displaymath}
    \frac{N_{CU}}{N_{U}} 
\end{displaymath}
where $N_{CU}$ represents, given a classifier and a document, the number of unique techniques correctly predicted by the classifier (i.e., regardless of how many times the techniques are detected across the sentences of the document), and $N_{U}$ indicates the total number of unique techniques predicted. Similarly, \textit{recall} was defined as:
\begin{displaymath}
    \frac{N_{CU}}{N_{GT}} 
\end{displaymath}
where $N_{CU}$ is the same number used to compute the precision, and $N_{GT}$ represents, for each document, the number of unique techniques in the ground truth. 




Figure \ref{fig:Document-eval} shows the results of our analysis. Given a document, we reported the value of F-measure for each classifier, for values of the threshold from $0.1$ to $0.8$, with a $0.1$ step. As expected, the results tend to be worse for higher values of the threshold, due to the fact that a smaller number of predictions is accepted. Indeed, the highest accuracy is reached with lower values of the threshold, with $0.2$ emerging as the best value for it. Intuitively, since the probability is split among all of 188 techniques, a $0.2$ probability for a given technique would be much higher than the other 187 ones.  
Overall, it is possible to notice that the DL models achieve better performance than the traditional ML models. The results of this analysis are lower than the sentence-level evaluation of the previous section. This can be traced back to the nature of the documents, which describe a large number of ATT\&CK techniques. Moreover, in a given document, the number of sentences that actually describe the techniques is usually lower than the number of non-pertinent ones. However, the average of F-measure achieved by the models is around 50\%: this is a good starting point to help a human operator who needs to analyze one or more documents, because the models will be capable of identifying at least half the techniques described in it.

\section{Lessons Learned}
\label{sec:lessons-learned}
In the experimental study, we investigated the use of machine learning to automatically classify unstructured CTI into the corresponding attack techniques from the MITRE ATT\&CK framework. The successes and open issues found in this study provide us with several lessons learned, as summarized in the following.


\vspace{3pt}
\noindent
\textbf{Applicability of ML for mapping unstructured CTI to attack techniques}. NLP to analyze unstructured text is a common procedure both in research and in the industry sector. The range of applicability of NLP is widening to different domains, including Cybersecurity in recent work. 
The experimental results provided by the first analysis with cross-validation support the idea that ML is indeed applicable on Cybersecurity documents. The classifiers achieved an accuracy (in terms of F-Measure) up to 72\%, even when challenged by a large set of classes (188 MITRE ATT\&CK techniques). This is a relevant result, since multi-class classification is a difficult task. 
Moreover, the ambiguity in natural language, as well as the ambiguity between classes (as discussed in Section \ref{sec:Analysis}) makes this problem even more challenging for ML algorithms, and poses challenges for their experimental evaluation, also.
Even if the practical adoption of these techniques seems feasible, further improvements are still required. 
In this work, we aim to lay the groundwork for developing new solutions for analyzing unstructured CTI, by providing new datasets, an evaluation methodology, and baseline results for comparisons. 

\vspace{3pt}
\noindent
\textbf{Selecting a ML algorithm}. From the initial analysis, we gained information on the performance of several different classifiers, spanning from traditional to deep learning-based ones. The best one has been SecureBERT, followed by MLP, SVM, Logistic Regression, and CNN. In particular, the analysis on several datasets of different sizes (i.e., TRAM and our own new dataset) suggests that deep learning-based models, such as CNN, have the potential to improve their accuracy when trained with larger datasets.
This finding is confirmed in the document-level analysis, where we found that traditional, well-known ML models, such as Logistic Regression and Naive Bayes, perform poorly compared to the others.

\vspace{3pt}
\noindent
\textbf{Applying classification on real-world documents}. Document-level evaluation provides a different point of view on the performance of ML algorithms. In this scenario, we deal with large unstructured texts that describe campaigns with several attack techniques, and that include additional non-pertinent sentences. Due to these additional challenges, the classification accuracy was generally lower compared to the previous sentence-level evaluation. 
Moreover, we found that several traditional ML algorithms were less accurate than the deep learning-based ones: this result points out that sentence-level evaluation is not sufficient for a thorough evaluation, since it can lead to over-optimistic results. Finally, our sensitivity analysis showed that the best accuracy is achieved when low values of the threshold are used to select output classes. Indeed, the best value for the threshold turned out to be $0.2$. Calibrating this threshold is important, since the classifier needs to deal with sentences that describe multiple attack techniques, and with other ones that do not include any technique.

\section{Related Work}
\label{sec:Related}
The automated analysis of Cyber Threat Intelligence is a recent and active research area in the cybersecurity field. 
\textit{TIM} \cite{you2022tim} is a framework that mines information about attack techniques from unstructured CTI using the Threat Context Enhanced Network (TCENet), a custom model developed for this task. It performs a coarse-grained analysis, by categorizing CTI into only 6 categories: Phishing, Scheduled Task/Job, Obfuscated Files or Information, Deobfuscate/Decode Files or Information, Collection, Application Layer Protocol. 
Ampel \emph{et al.} \cite{ampel2021linking} developed a model to analyze Common Vulnerabilities and Exposures (CVEs) and to label them with one of 10 MITRE ATT\&CK tactics (Defense Evasion, Discovery, Privilege Escalation, Collection, Lateral Movement, Impact, Credential Access, Initial Access, Exfiltration, Execution). 
\textit{EXTRACTOR} \cite{satvat2021extractor} is a tool for analyzing sentences to extract information on attack behavior. In particular, EXTRACTOR identifies system entity names (e.g., file or process names, IP, and registry keys) and actions (read, write send, receive, connect, fork) described in the sentences. This approach performs a low-level analysis, without abstracting the actions to ATT\&CK techniques. 
\textit{TTPDrill} \cite{husari2017ttpdrill} is a system that extracts threat actions from unstructured sources of CTI. It leverages a dedicated ontology to fit the mined malicious actions, by combining NLP and Information Retrieval (IR) techniques. The scope of this tool is limited to the identification of a small set of threat actions, such as data exfiltration, add registry, and DLL injection. 
\textit{ActionMiner} \cite{husari2018using} is another framework for the extraction of threat actions from CTI documents. It uses entropy and mutual information to identify low-level actions, which are represented as verb-object couples. These threat actions are not mapped to the techniques in the MITRE ATT\&CK taxonomy.
Our work differs from these previous ones. Most importantly, our approach classifies CTI by mapping it into all of the 188 ATT\&CK techniques. This fine-grain analysis is particularly important in some scenarios, including the planning of adversary emulation activities, which need to match the specific actions of threat actors to be realistic. Moreover, in this work, we performed a more extensive study on the automatic classification of CTI, considering several datasets and ML models.

\section{Threats to validity}

\vspace{3pt}
\noindent
\textbf{Documents.} To build the datasets, we selected threat and incident reports publicly available online. As a consequence, the choice of the CTI for the dataset could be affected by our selection. To mitigate this issue, we focused on documents gathered by MITRE for the ATT\&CK framework.

\vspace{3pt}
\noindent
\textbf{Models.} For the classification of attack techniques, we analyzed several traditional ML and Deep Learning models. We selected the most popular models for NLP tasks, choosing both traditional and recent ones. We also considered a Transformer-based model, \textit{SecureBERT}, so as to include the most recent advancements in NLP research. To the best of our knowledge, our study covers the largest number of ML models in this field.

\vspace{3pt}
\noindent
\textbf{Metrics.} To evaluate the performance of the classification models, we adopted the traditional metrics for multi-class classification tasks. In particular, we decide to consider the \textit{F1-Score} instead of the \textit{accuracy}, to take into account the imbalance of our datasets and the distribution of classes. We also selected the \textit{top K accuracy}, with K set to 3, which indicates if the correct class is in the top 3 predictions.

\section{Conclusion and Future Work}
\label{sec:Conclusion}
In this work, we presented an experimental study on the automatic mapping of Cyber Threat Intelligence into attack techniques. In particular, we focused on the well-known MITRE ATT\&CK taxonomy to define classes to be mapped to CTI. Moreover, we adopted the knowledge base of MITRE ATT\&CK to build two datasets of CTI descriptions in natural language: the first one labeled at sentence-level, the other at document-level, with the corresponding attack techniques. We tested several state-of-the-art classification models, which achieved promising results at mapping attack techniques. The lessons learned from this analysis point out the strengths and limitations of the aforementioned ML models, and set directions for future work. 

In addition to improving the mapping of CTI to ATT\&CK techniques, we foresee future work on integrating this approach into offensive security practices. In particular, adversary emulation can benefit from this work, in order to enable organizations to emulate the actions of a given threat actor, according to CTI collected from field. The mapped CTI can potentially be used to configure an automated tool, in order to emulate the specific attack techniques used by a threat actor. This, in turn, would significantly reduce the cost and effort needed to perform security assessments and training exercises.

\section*{Acknowledgments}
\label{sec:Ack}
This work has been partially supported by the Programme F.R.A. of Università degli Studi di Napoli Federico II (project OSTAGE), and by the Italian Ministry of University and Research (MUR) under the programme ``PON Ricerca e Innovazione 2014-2020 – Dottorati innovativi con caratterizzazione industriale''.

\bibliographystyle{plain}
\bibliography{biblio}

\end{document}